\documentclass[aip,jmp,amsmath,amssymb,groupaddresses,superscriptaddress
]{revtex4}

\usepackage{graphicx}%
\usepackage{dcolumn}%
\usepackage{bm}%
\usepackage{color}
\usepackage{enumerate} 

\def\e{\mathrm{e}}
\def\ii{\mathrm{i}}
\def\d{\mathrm{d}}

\newtheorem{definition}{Definition}
\newtheorem{theorem}{Theorem}
\newtheorem{lemma}[theorem]{Lemma}
\newtheorem{proposition}{Proposition}

\begin{document}

\title{Invariant measures on multimode quantum  Gaussian states}

\author{C. Lupo}
\affiliation{School of Science and Technology, Universit\`a di Camerino, I-62032 Camerino, Italy.}

\author{S. Mancini}
\affiliation{School of Science and Technology, Universit\`a di Camerino, I-62032 Camerino, Italy.}
\affiliation{Istituto Nazionale di Fisica Nucleare, Sezione di Perugia, I-06123 Perugia, Italy.}

\author{A. De Pasquale}
\affiliation{NEST, Scuola Normale Superiore and Istituto Nanoscienze-CNR, I-56126 Pisa, Italy.}

\author{P. Facchi}
\affiliation{Dipartimento di Matematica and MECENAS, Universit\`a di Bari, I-70125  Bari, Italy.}
\affiliation{Istituto Nazionale di Fisica Nucleare, Sezione di Bari, I-70126 Bari, Italy.}

\author{G. Florio}
\affiliation{Museo Storico della Fisica e Centro Studi e Ricerche ÒEnrico FermiÓ, Piazza del Viminale 1, I-00184 Roma, Italy.}
\affiliation{Dipartimento di Fisica and MECENAS, Universit\`a di Bari, I-70126  Bari, Italy.}
\affiliation{Istituto Nazionale di Fisica Nucleare, Sezione di Bari, I-70126 Bari, Italy.}

\author{S. Pascazio}
\affiliation{Dipartimento di Fisica and MECENAS, Universit\`a di Bari, I-70126  Bari, Italy.}
\affiliation{Istituto Nazionale di Fisica Nucleare, Sezione di Bari, I-70126 Bari, Italy.}

\date{\today}

\begin{abstract}
We derive the invariant measure on the manifold of multimode quantum Gaussian states,
induced by the Haar measure on the group of Gaussian unitary transformations.
To this end, by introducing a bipartition of the system in two disjoint subsystems,
we use a parameterization highlighting the role of nonlocal degrees of freedom
--- the symplectic eigenvalues --- which characterize quantum entanglement across
the given bipartition.
A finite measure is then obtained by imposing a physically motivated energy constraint. 
By averaging over the local degrees of freedom we finally derive the invariant
distribution of the symplectic eigenvalues in some cases of particular interest 
for applications in quantum optics and quantum information.
\end{abstract}

\keywords{Quantum Gaussian states, Haar measures, Entanglement characterization}

\maketitle

\section{Introduction}

With the advent of quantum information theory \cite{nielsenchuang} the notion of entanglement has assumed a prominent role \cite{entanglement}.
Bipartite entanglement is the simplest to consider. It can be quantitatively characterized in terms of 
several (though physically equivalent) measures in the case of small quantum systems (such as a pair of qubits).  
The problem becomes more complicated for larger systems where, also due to the exponentially increasing 
complexity of the Hilbert space resulting from multiple constituents, a theoretical characterization 
turns out to be a daunting task. 

A promising approach towards such a characterization relies on statistical properties of 
entanglement when the states of the system are assumed to be distributed according to a 
suitable probability measure. In this way the problem is simplified by restricting attention 
on the `typical' (most likely) features of the entanglement.
For finite dimensional quantum systems a natural and unbiased measure on pure states stems 
from the Haar measure of the unitary group, whose elements allow to retrieve any pure state 
when applied to another fixed pure state. On this basis typical entanglement can be addressed and analyzed.

More specifically, let us consider a collection of $n$ $d$-dimensional quantum systems described by a Hilbert space 
$\mathcal{H} \simeq \mathbb{C}^{d^n}$, and a bipartition $\mathcal{H}=\mathcal{H}_A \otimes \mathcal{H}_B$
into two disjoint sets of respectively $n_A$ and $n_B$ elementary systems, with $\mathcal{H}_A \simeq  \mathbb{C}^{d^{n_A}}$, 
$\mathcal{H}_B  \simeq  \mathbb{C}^{d^{n_B}}$ and $n_A+n_B=n$.
Putting $\mathcal{N}_A=d^{n_A}$, $\mathcal{N}_B=d^{n_B}$, $\mathcal{N}=d^n$ and assuming without loss of generality $n_A \leq n_B$,
a normalized vector $|\psi\rangle \in \mathcal{H}$, representing a pure state of the system,
can always be put in the following normal form (Schmidt decomposition)~\cite{Peres}: 
\begin{equation}
|\psi\rangle = \sum_{j=1}^{\mathcal{N}_A} \sqrt{p_j} \, |j\rangle_A \otimes |j\rangle_B \, .
\end{equation}
Here $\{ |j\rangle_A \}_{j=1,\dots,\mathcal{N}_A}$, $\{ |j\rangle_B \}_{j=1,\dots,\mathcal{N}_B}$
are orthonormal systems of the local Hilbert spaces $\mathcal{H}_A$ and $\mathcal{H}_B$, respectively. 
The components of the probability vector $p=(p_1,\dots,p_{\mathcal{N}_A})\in \Delta_{\mathcal{N}_A-1}$, where 
$\Delta_{\mathcal{N}_A-1}$ is the $(\mathcal{N}_A-1)$-simplex ($p_k\ge0$, $\sum p_k=1$), are known as the Schmidt coefficients. 
They completely characterize the nonlocal features of bipartite entanglement in pure states~\cite{Vidal}.
Then, accordingly to what has been stated above, the statistical properties of entanglement can be found once
the distribution of the Schmidt coefficients has been defined.
A suitable distribution is the one induced by the Haar measure on the $\mathcal{N}$-dimensional
unitary group $\mathrm{U}(\mathcal{N})$, that reads ~\cite{Lloyd,Lubkin}
\begin{equation}\label{DSchmidt}
\d\mu_S (p) = P(p) \, \d p = C_{\mathcal{N},\mathcal{N}_A} \prod_{h>k=1}^{\mathcal{N}_A} \left( p_h - p_k \right)^2 \prod_{j=1}^{\mathcal{N}_A} p_j^{\mathcal{N}_B-\mathcal{N}_A} \d p \, ,
\end{equation}
with $\d p$ the Lebesgue  measure on $\Delta_{\mathcal{N}_A-1}$, and $C_{\mathcal{N},\mathcal{N}_A}$ a normalization factor. 
The probability measure $P(p)$ is by construction invariant under the action of the unitary group 
$\mathrm{U}(\mathcal{N})$.
This expression has been the starting point of several studies concerning the
typicality of entanglement, see e.g.\ Refs.~\cite{Page,Giraud,PTE1,PTE2,mixmatrix,Nadal}. 

For any given number of elementary systems $n$, $n_A$ and $n_B$, the invariant distribution~(\ref{DSchmidt}) 
is not well defined in the case of elementary Hilbert spaces with infinite dimensions, i.e., in the limit $d \to \infty$.
This makes problematic the characterization of entanglement in continuous variable quantum systems,
i.e., quantum systems described by pairs of canonically conjugated observables with continuous spectra, 
for which $\mathcal{H} = L^2(\mathbb{R})^{\otimes n} \simeq L^2(\mathbb{R}^n)$.
However, due to the relevance of such systems \cite{BvL,FOP,review2} a solution of this problem urges.
To this end we shall consider a specific finite-dimensional, yet unbounded, manifold 
$\mathcal{G}\subset L^2(\mathbb{R})^{\otimes n}$ of pure states, known as Gaussian states~\cite{Simon87,Simon88,Simon94,BvL,FOP,review2}.

Entanglement typicality in multimode Gaussian states has been the subject of few 
investigations~\cite{Adesso-PRL,Serafini1,Serafini2,Serafini_Adesso_JPA}, whose starting points were the
probability measures induced by microcanonical-like and canonical-like ensembles.
Here, we follow a different route and derive an invariant probability measure induced
by the 'most unbiased' measure, i.e., the Haar measure on the group of
Gaussian unitary transformations, whose orbit passing through the vacuum state is
the manifold of Gaussian states (this is the content of Theorem \ref{th1} of Section \ref{Sec:1stM} 
and parallels Eq.~\eqref{DSchmidt}).
In the same fashion we derive the invariant measure on the group of $n$-mode Gaussian 
unitary transformations (this is the content of Theorem \ref{theoremunitmeasure} of Section \ref{Sec:1stM}).
Clearly, although the Haar measure can be considered the `most unbiased' measure, other choices
are possible and have been considered in literature~\cite{Adesso-PRL,Serafini1,Serafini2,Serafini_Adesso_JPA} 
as starting point to investigate entanglement typicality in multimode Gaussian states.

Furthermore, in order to obtain a measure which is finite and motivated by physical applications, 
in Section \ref{Sec:Ex} we introduce a suitable effective cutoff by means 
of an energy constraint. That enables us to normalize the invariant measure, then to compute the 
average over the local degrees of freedom and finally to derive the probability distribution of 
the nonlocal parameters (symplectic eigenvalues) characterizing bipartite entanglement in Gaussian states. 
Among other examples, we consider a specific submanifold of Gaussian states which is relevant for applications 
in the domain of quantum optics, namely the submanifold of states generated by nonlinear optical parametric processes. 
Indeed, the latter induce a natural bipartition of the set of modes into two disjoint subsets, respectively
called {\it signal} and {\it idler} modes, which breaks the symmetry under mode permutations.

The present study provides tools for attacking the theory of continuous variable entanglement 
which contains a variety of (locally) inequivalent classes. 
In fact, restricting statements to the `typical entanglement' would allow to overcome several complications. 
That was already pointed out in Ref.~\onlinecite{Hayden}, for finite dimensional systems, by looking at `concentration of measure' 
around the average of the entanglement probability distribution with increasing $n$. 
Furthermore, our study can give insights into the complexity of entangling quantum 
circuits with continuous variable gates. Remarkably, in the finite dimensional case Eq.~\eqref{DSchmidt} 
leads to the fact that a circuit of elementary quantum gates is able to maximally entangle 
a separable state within an arbitrary accuracy by a number of gates that grows only polynomially 
in the number of qubits of the register \cite{ODP_PRL,DOP_JMP}.
Finally, since the statistical properties of entanglement can be derived by introducing a partition 
function with a fictitious temperature, the present study can help in singling out possible phase 
transitions between different regions of entanglement corresponding to different ranges of temperature, 
similarly to what occurs in the finite dimensional case \cite{PTE1,PTE2}.

\section{Definitions and main results}\label{Sec:defmain}

\subsection{Multimode Gaussian States}\label{Sec:MMGS}

In this SubSection we briefly recall few basic definitions and properties of Gaussian states. 
For a comprehensive review of their mathematical features and physical relevance we refer to~\cite{Simon87,Simon88,Simon94,BvL,review2}.

\begin{definition}[Continuous variable quantum systems]\label{def1}
\emph{A continuous variable (CV) quantum system} is defined by a collection of $n$ bosonic modes in the Hilbert space $L^2(\mathbb{R}^n)$ with canonical annihilation and creation operators 
\begin{equation}
a_k, a_k^\dag \, , \quad k=1 , \dots , n ,
\end{equation}
defined on the common dense domain $\mathcal{S}(\mathbb{R}^n)$, the Schwartz space of the functions of rapid decrease,  and acting as
\begin{equation}
a_k \psi (x) = 2^{-1/2}(x_k + \partial/ \partial x_k)\psi(x) \, , \qquad 
a_k^\dagger \psi(x) = 2^{-1/2}(x_k - \partial/ \partial x_k) \psi(x) \, ,
\end{equation}
for any $\psi \in \mathcal{S}(\mathbb{R}^n)$. They obey
the canonical commutation relations
\begin{equation}
[a_h, a_k^\dag] = \delta_{hk} \, , \quad [ a_h, a_k ] = [ a_h^\dag, a_k^\dag ] = 0 \, , \quad\forall\, h,k=1\dots n .
\end{equation}
We define the $n$-mode \emph{vacuum state} $|0\rangle
\in \mathcal{S}(\mathbb{R}^n)$, as the  unit vector satisfying
\begin{equation}
a_h |0\rangle = 0 \, ,
\end{equation}
and given by 
\begin{equation}\label{eq:vacuum} 
\phi_0(x) = \pi^{-n/4} \e^{-|x|^2/2} \, .
\end{equation}
It generates the  \emph{$n$-mode number states} with occupation numbers $(j_1, j_2, \dots, j_n) \in \mathbb{N}^n$ by 
\begin{equation}\label{eq:Hermite}
| j_1, j_2, \dots,j_n \rangle := \prod_{k=1}^n (j_k!)^{-1/2} (a_k^\dag)^{j_k} |0\rangle \, .
\end{equation}
\end{definition}

Notice that the $n$-mode number states $| j_1, j_2, \dots,j_n \rangle\in \mathcal{S}(\mathbb{R}^n)$ are just 
tensor products  of Hermite functions (Hermite polynomials times a Gaussian), which form an orthonormal basis 
for $L^2(\mathbb{R}^n)$.
 In the following we will study the action of multimode Gaussian unitaries, namely those unitary operators 
generated by quadratic self-adjoint polynomials of the canonical operators.

\begin{definition}[Gaussian unitary transformations]\label{def:quadHam}
Let us consider the family of \emph{quadratic Hamiltonians}, as the family of symmetric operators 
on the common domain $\mathcal{S}(\mathbb{R}^n)$, that are quadratic in the canonical operators. Namely,
\begin{equation}\label{GUnitary}
H_G(\theta, \alpha) =H_0+H_1+H_2 \, , 
\end{equation}
where
\begin{eqnarray}
H_0 &=& \theta \, , \label{eq:constham} \\
H_1 &=& \ii \sum_{k=1}^n \left( \xi_k a^\dag_k - \xi_k^* a_k \right) \, , \label{eq:linearham} \\
H_2 &=&\sum_{h,k=1}^n \left( M_{hk} a_h a_k + M_{hk}^* a_h^\dag a_k^\dag + N_{hk} a_h^\dag a_k \right) \, , \label{eq:quadraticham}
\end{eqnarray}
with $\theta \in \mathbb{R}$, $\alpha = (\xi, M,N)$, where $\xi \in \mathbb{C}^n$, $M$ is a complex and symmetric matrix and 
$N$ is a Hermitian matrix.
We define as \emph{multimode Gaussian unitary} (or \emph{$n$-mode Gaussian unitary}) the exponential of $H_G$:
\begin{equation}\label{def:gaussianunitary}
\mathcal{U}_G(\theta, \alpha) = \e^{-\ii H_G(\theta, \alpha)} \, .
\end{equation}
\end{definition}
The set of $n$-mode Gaussian unitaries is a group, denoted as the {\it group of $n$-mode Gaussian unitaries}.
The quadratic Hamiltonians of the form (\ref{GUnitary}) constitute the Lie algebra of this group.
It is well known that the group of $n$-mode Gaussian unitaries is a
projective representation of the inhomogeneous symplectic group $\mathrm{ISp}(2n,\mathbb{R})$,
with the quadratic terms~(\ref{eq:quadraticham}) corresponding to the homogeneous 
subgroup $\mathrm{Sp}(2n,\mathbb{R})$, and the linear terms~(\ref{eq:linearham}) to the subgroup of translations
$T_{2n}$, isomorphic to $\mathbb{C}^n$ (in quantum optics $H_1(\xi)$ represents the generator of 
the {\it displacement operator} and $\xi$ is the amount of displacement in the optical phase).
We hence denote as {\it homogeneous Gaussian unitaries} the Gaussian unitary transformations obtained
by putting $\xi=0$.
Finally, notice that $\mathcal{S}(\mathbb{R}^n)$ contains the linear span of the Hermite functions, 
which is a dense set of analytic vectors for any $H_G$.
Therefore, by Nelson's analytic vector theorem, $\mathcal{S}(\mathbb{R}^n)$ is a common domain of 
essential self-adjointness of the quadratic Hamiltonians \cite{ReedSimon}.

Our analysis focuses on the manifold of \emph{pure Gaussian states} $\mathcal{G}\subset\mathcal{H}= L^2(\mathbb{R}^n)$ 
defined as the orbit of the action of the group of Gaussian unitary operators on the vacuum state:
\begin{definition}[Gaussian pure states]
We call any vector state of the form 
\begin{equation}
\psi_G = \mathcal{U}_G |0\rangle = \e^{-\ii H_G} |0\rangle \, ,
\end{equation}
for some multimode Gaussian unitary $\mathcal{U}_G$, a \emph{Gaussian pure state}. 
\end{definition}
The orbit of the subgroup of homogeneous Gaussian unitaries is a submanifold of Gaussian states,
that we refer to as the submanifold of {\it homogeneous Gaussian states}. 
Homogeneous Gaussian states are characterized by the conditions
\begin{equation}
\langle \psi_G | a_k  \psi_G \rangle = \langle \psi_G | a_k^\dag  \psi_G \rangle = 0 \, , \qquad k=1,\dots,n \, .
\end{equation}
According to the definition of Perelomov~\cite{Perelomov}, Gaussian states are coherent states for
$\mathrm{ISp}(2n,\mathbb{R})$, and the homogenous Gaussian states are coherent
states for $\mathrm{Sp}(2n,\mathbb{R})$.

Let us hence consider a bipartition of the system into two disjoint subsets, denoted $A$ and $B$, 
of the canonical modes,
\begin{eqnarray}
{a}_k, {a}_k^\dag \, , \qquad {k=1,\dots, n_A} \, , \\ 
{b}_k, {b}_k^\dag \, , \qquad {k=1,\dots, n_B} \, , 
\end{eqnarray}
with $n_A + n_B = n$. This naturally induces on the total Hilbert space 
the bipartition into two Hilbert spaces, 
$\mathcal{H}= \mathcal{H}_A \otimes \mathcal{H}_B= L^2(\mathbb{R}^{n_A}) \otimes L^2(\mathbb{R}^{n_B})$.
Without loss of generality we assume $n_A \leq n_B$.
For Gaussian pure states of a bipartite system the following Proposition holds:
\begin{proposition}[Ref.~\onlinecite{canonical}]\label{prop:BR}
By acting with local Gaussian unitary
transformations $\mathcal{U}_G^A$ and $\mathcal{U}_G^B$ on subsystems $A$ and $B$, any Gaussian pure state $\psi_G$  
of  $n_A + n_B$ modes can be put into the canonical form
\begin{eqnarray} \label{CVS}
\psi_G^c  = \mathcal{U}_G^A \otimes \mathcal{U}_G^B\, \psi_G = \prod_{k=1}^{n_A} \exp{\left[ r_k ({a}_k {b}_k -  {a}_k^\dag {b}_k^\dag) \right]} |0\rangle \, , 
\end{eqnarray}
where $r_1, r_2, \ldots, r_{n_A}$ are  nonnegative parameters (unique up to permutations)
associated to the state $\psi_G$.
\end{proposition}

In order to clarify our notation, let us  consider the case $n=2$, with $n_A=n_B=1$. 
Then any two mode Gaussian state, by local Gaussian unitaries, can be put in the canonical form
\begin{equation}\label{TMSV}
 |\textrm{TMSV}\rangle = \exp{\left[ r ({a} {b} -  {a}^\dag {b}^\dag) \right]} |0\rangle \, , 
\end{equation}
which is the so-called {\it two-mode squeezed vacuum} (or {\it twin-beam state}),
a state of utmost importance in quantum optics and in CV implementations of quantum information
processing~\cite{BvL}. 

By expanding the exponential in Eq.~(\ref{TMSV}) one gets
\begin{equation}
|\textrm{TMSV}\rangle = \sqrt{\frac{2}{\nu+1}} \sum_{j=0}^{\infty} \left(\frac{\nu-1}{\nu+1}\right)^{j/2}  |j, j \rangle \, ,
\end{equation}
where $\nu = \cosh{2r}$.
By taking the partial trace over one of the two modes, one gets a thermal-like reduced state
\begin{equation}
\rho_A = \rho_B = \frac{2}{\nu+1} \sum_{j=0}^\infty \left(\frac{\nu-1}{\nu+1}\right)^j |j\rangle \langle j| \, . 
\end{equation}
Thus, the entanglement of the two-mode squeezed vacuum is characterized 
by the single parameter $\nu\in [1,+\infty)$. 

Coming back to the multimode case, by means of the canonical decomposition in Eq.~(\ref{CVS}),
we see that it is possible to write any Gaussian pure state of $n_A+n_B$ modes
as the direct product of $n_A$ two-mode squeezed vacua and $n_B-n_A$ vacua.
\begin{definition}[Symplectic eigenvalues]
Given a Gaussian pure state $\psi_G \in \mathcal{G}$, and a bipartition into $n_A + n_B$ modes, 
with $n_A \leq n_B$, the \emph{symplectic eigenvalues} associated to the given bipartion are defined as
\begin{equation}\label{eq:defseign}
\nu_k := \cosh{2r_k} \,, \qquad k=1,\dots, n_A \, ,
\end{equation} 
where the parameters $r_k$ are given by the canonical form~(\ref{CVS}) of $\psi_G$.
\end{definition}

All the entanglement properties of a $n$-mode Gaussian pure state with respect to the given
bipartition are uniquely determined by the $n_A$ symplectic eigenvalues, depending on the 
{\it nonlocal} parameters $r_k$. They are invariant under local unitary transformations and can be explicitly 
computed by symplectic diagonalization~\cite{canonical,HolevoWerner}.

\subsection{Main results}\label{Sec:1stM}

The manifold $\mathcal{G}$ of Gaussian states is a locally compact orbit of the group
of Gaussian unitary transformations. 
Indeeed, through the parameterization of Def.~\ref{def:quadHam} it is locally homeomeorphic to a finite dimensional Euclidean space.
Therefore, there exists a unique invariant measure on the manifold of Gaussian states 
induced by the left Haar measure on the group of Gaussian unitaries.
Our main result is the derivation of the explicit form of the invariant measure on Gaussian
states. 
In particular, being interested in the characterization of entanglement, we use the latter 
to compute the probability measure on the symplectic eigenvalues in several settings of
physical relevance.

The following Theorem is the main result of this paper:
\begin{theorem}\label{th1} 
The invariant measure on the manifold of Gaussian pure states has the form
\begin{equation}\label{invm}
\d\mu_G = K_{n,n_A} 
\prod_{h>k=1}^{n_A} \left( \nu_h^2 - \nu_k^2 \right)^2 \prod_{j=1}^{n_A} \nu_j^2 (\nu_j^2-1)^{n_B-n_A} \, \d\nu \, 
\d\mu_A(\alpha_A) \, \d\mu_B(\alpha_B) \, \d\theta \, ,
\end{equation}
where: $\nu=(\nu_1, \dots, \nu_{n_A})$ are the symplectic eigenvalues and $\d\nu=\prod_{h=1}^{n_A} \d\nu_h$; 
$\d\mu_A(\alpha_A)$, $\d\mu_B(\alpha_B)$ with $\alpha=(\xi,M,N)$ of Def.~\ref{def:quadHam}, 
are the volume forms of the local degrees of freedom $\alpha_A$ and $\alpha_B$ induced by the Haar 
measure on the local subgroups of Gaussian unitaries acting on subsystems $A$ and $B$ respectively;
$\d\theta$ corresponds to the scalar term in Definition \ref{def:quadHam}; and $K_{n,n_A}$ is a normalization factor.
\end{theorem}
The proof of Theorem \ref{th1} is given in Section \ref{proof}.

This expression for the invariant measure on the manifold of $n$-mode Gaussian states 
makes use of a parameterization which highlights the symplectic eigenvalues as signatures of 
the entanglement across a given bipartition of the system.

\subsubsection{Haar measure on Gaussian unitary transformations}\label{sssec:GU}

The expression in Eq.~(\ref{invm}) can be used to derive the distribution of the symplectic
eigenvalues by integrating over the parameter $\theta$ and over the local degrees of freedom 
$\alpha_A$, $\alpha_B$.

It is hence worth deriving an explicit expression for the Haar measure on the group of
$n$-mode Gaussian unitary transformations. 
For the sake of conciseness here we restrict to the subgroup of homogenous Gaussian 
unitary transformations --- obtained by setting $\xi=0$ in Eq.~(\ref{GUnitary}) ---,
the extension to the non-homogenous group being straightforward.

A parameterization of the subgroup of homogenous Gaussian unitary transformations 
can be obtained from the Euler decomposition:
\begin{equation}\label{gaussunitary2}
\mathcal{U}_G =\e^{-\ii \theta} \exp{\left( -\ii \sum_{i,j=1}^n T_{ij} a_i^\dag a_j \right)} \exp{\left( \sum_{k=1}^n s_k a_k^2 - s_k (a_k^\dag)^2 \right)}
\exp{\left( -\ii \sum_{i,j=1}^n T'_{ij} a_i^\dag a_j \right)} \, ,
\end{equation}
where $\theta \in \mathbb{R}$, $T$ and $T'$ are Hermitian matrices, and $s_1, s_2, \dots, s_n$ 
are real and non-negative parameters. 
The Gaussian unitaries of the form $\mathcal{U}_{\bar{G}}=\exp{\left( -\ii \sum_{i,j=1}^n T_{ij} a_i^\dag a_j \right)}$,
with $T$ Hermitian, define a representation of the group $\mathrm{U}(n)$, see e.g.\ Ref.~\onlinecite{Aniello}, 
and their action on the canonical operators reads
\begin{equation}\label{JS-id}
\mathcal{U}_{\bar{G}}^\dag \, a_k \, \mathcal{U}_{\bar{G}} = \sum_{h=1}^n U_{kh} \, a_h \, ,
\end{equation}
where $U = \e^{-\ii T} \in \mathrm{U}(n)$.
Furthermore one has,
\begin{equation}\label{localsq}
\exp{\bigg(-\sum_{j=1}^n s_j a_j^2 - s_j (a_j^\dag)^2\bigg) } \, a_k \, \exp{\bigg(\sum_{j=1}^n s_j a_j^2 - s_j (a_j^\dag)^2\bigg)} =
\cosh{(2s_k)} \, a_k - \sinh{(2s_k)} \, a_k^\dag \, .
\end{equation}

 The following Theorem holds:
\begin{theorem}\label{theoremunitmeasure}
The Haar invariant measure on the group of $n$-mode homogeneous Gaussian 
unitaries, parameterized according to Eq.~(\ref{gaussunitary2}), is given by
\begin{equation}
\d\mu(\mathcal{U}_G) = K_n \prod_{h<k=1}^n | \lambda_h - \lambda_k | \; \d\lambda \, \d\mu_H(U)\,  \d\mu_H(U') \, ,
\end{equation}
where $\mu_H$ denotes the Haar invariant measure on the unitary group $\mathrm{U}(n)$, $U = \exp(-\ii T)$, $U' = \exp(-\ii T')$,
$\lambda = (\lambda_1, \lambda_2, \dots, \lambda_n)$, $\lambda_k=\cosh{2s_k}$, $\d\lambda = \prod_{k=1}^n \d\lambda_k$, and $K_n$ 
is a normalization factor.
\end{theorem}
The proof of Theorem \ref{theoremunitmeasure} is given in Section \ref{measure2}.

\subsubsection{Distributions of the symplectic eigenvalues}\label{sssec:PP}

Section \ref{Sec:Ex} presents several examples of distributions of the symplectic eigenvalues.
Since the manifold of Gaussian states --- although locally compact --- is not compact, the 
invariant measures as well as the distribution of the symplectic eigenvalues cannot be globally
normalized. To circumvent this problem one may introduce a suitable effective cutoff.
Motivated by general physical considerations we hence introduce an effective cutoff by imposing
a finite value of the mean energy of each of the subsystems $A$ and $B$. This allows us to derive
a finite measure on the manifold of Gaussian states and, after integration on the local degrees of
freedom, a normalized distribution of the symplectic eigenvalues. Explicit expressions are derived
for the case of a CV quantum systems composed of $1+1$ and $2+2$ modes.

Finally, in SubSection \ref{Sec:subM}, Theorem \ref{theoremmeasure3}, we derive the measure on 
a submanifold of Gaussian states which is of special interest in view of physical applications. 
This submanifold of states arises in nonlinear optics when the Hamiltonian governing the physical 
process induces a natural bipartition of the set of modes into two disjoint subsets, respectively 
called the {\it signal} and {\it idler} modes, which breaks the symmetry under permutations of the modes.
The mean energy takes a particularly simple expression for these states, allowing us to derive 
the corresponding symplectic eigenvalue distribution under constrained mean energy.

\section{Applications and examples}\label{Sec:Ex}

\subsection{A finite measure from an energy constraint}

The manifold of Gaussian states is not compact, implying that the invariant measure
cannot be globally normalized.
As a consequence, all the statistical moments of the symplectic eigenvalues diverge. 
This also yields unbounded statistical moments of entropies of entanglement~\cite{Agarwal}. 
The same holds true for other entanglement quantifiers, as for instance the logarithmic negativity~\cite{VW,FOP} 
and the coherent information~\cite{nielsen,HolevoWerner}, which are increasing functions of the symplectic eigenvalues.
In order to avoid unphysical results we impose a suitable constraint, yielding an 
effective cutoff on the unbounded manifold of Gaussian states, see also Refs.~\onlinecite{HolevoWerner}, \onlinecite{GMMES1}, \onlinecite{GMMES2}.
There are clearly many inequivalent ways in which such a cutoff can be introduced that will lead
to different distributions of the symplectic eigenvalues.
Here we choose to constrain the mean value of the energy in each subsystem.
Such a choice is motivated by physical realizations of Gaussian states, especially in the field of quantum optics, 
in which the mean energy is a crucial quantity \cite{BvL,review2}.
\begin{definition}[Mean energy]
Given a Gaussian pure state $\psi_G$ and a bipartition  of $n$ harmonic oscillators in two subsystems $A$ and $B$, all with frequency $\omega_k=1$, $k=1,\ldots,n$ and $\hbar=1$, the mean energy of each subsystem is defined as
\begin{eqnarray}
\mathcal{E}_A & = & \frac{1}{2} \sum_{k=1}^{n_A} \langle \psi_G | ( {a}^\dag_k {a}_k + a_{k} {a}^\dag_k ) \psi_G \rangle \, ,  \\
\mathcal{E}_B & = & \frac{1}{2} \sum_{k=1}^{n_B} \langle \psi_G | ( {b}^\dag_k {b}_k + {b}_k {b}^\dag_k ) \psi_G \rangle \, . 
\end{eqnarray}
\end{definition}
The mean energies take values in the interval $[1/2,+\infty)$. 

In the following we will consider the submanifold of homogeneous Gaussian states and restrict to 
the case of balanced bipartitions with $n_A = n_B = n/2$ ($n$ even). The motivation of this choice is based on the fact that the use of balanced bipartitions maximizes the amount of information when studying the properties of the symplectic eigenvalues and, therefore, of entanglement. However, the extension to non-homogeneous Gaussian states and/or unbalanced bipartition is straightforward. 
The subsystems mean energies are the quantities that we will consider fixed 
(and finite) on a physical basis. 
The following Lemma gives a convenient expression for $\mathcal{E}_A$ and $\mathcal{E}_B$:
\begin{lemma}\label{theoremmeanen}
The mean energies of subsystems A and B, when calculated on homogeneous Gaussian states
and for $n_A=n_B=n/2$, have the form
\begin{eqnarray}
\mathcal{E}_A & = & \frac{1}{2} \sum_{h,k=1}^{n/2} |{U_A}_{hk}|^2 \, {\lambda_A}_h \, \nu_k \, , \label{conA} \\
\mathcal{E}_B & = & \frac{1}{2} \sum_{h,k=1}^{n/2} |{U_B}_{hk}|^2 \, {\lambda_B}_h \, \nu_k  \, , \label{conB}
\end{eqnarray}
where ${U_A}$ and ${U_B}$ are unitary matrices of $\mathrm{U}(n/2)$, $\nu$ are the symplectic eigenvalues defined in Eq.~(\ref{eq:defseign}), 
and $\lambda_A=( {\lambda_A}_1, {\lambda_A}_2, \dots, {\lambda_A}_{n/2} )$ , $\lambda_B = ( {\lambda_B}_1, {\lambda_B}_2, \dots, {\lambda_B}_{n/2} )$ 
are the parameters defined in Theorem \ref{theoremunitmeasure}. 
\end{lemma}
The proof of the Lemma is given in Section~\ref{App-energy}.
Notice that the mean energies are functions of both the nonlocal parameters $\nu$ 
and the local ones, $U_A$, $\lambda_A$, $U_B$, $\lambda_B$.

By imposing the constraints \eqref{conA}, \eqref{conB} on the local mean energies, we get an expression for the probability measure of the symplectic eigenvalues:
\begin{eqnarray}
& & \d\mu(\nu | \mathcal{E}_A,\mathcal{E}_B) = P(\nu | {\mathcal{E}_A,\mathcal{E}_B}) \, \d\nu \nonumber \\
& & = \int \delta\bigg( \mathcal{E}_A -\frac{1}{2} \sum_{h,k=1}^{n/2} |{U_A}_{hk}|^2 \, {\lambda_A}_h \, \nu_k \bigg) \,
\delta\bigg( \mathcal{E}_B -\frac{1}{2} \sum_{h,k=1}^{n/2} |{U_B}_{hk}|^2 \, {\lambda_B}_h \, \nu_k \bigg) \,  {\d\mu_G} \, , 
\end{eqnarray}
where $\delta$ denotes the Dirac delta function.
This integral can be evaluated by using the invariant measure of Eq.~\eqref{invm}
after making the formal substitutions of $\d\mu(\alpha_A)$, $\d\mu(\alpha_B)$
with $\d\mu(\mathcal{U}_G^A)$, $\d\mu(\mathcal{U}_G^B)$, 
the latter being the Haar invariant measures on the local groups of Gaussian
unitaries. The explicit form of the Haar measures $\d\mu(\mathcal{U}_G^A)$, $\d\mu(\mathcal{U}_G^B)$ 
is given by Theorem~\ref{theoremunitmeasure}.
After integrating we obtain
\begin{equation}\label{EConstraint}
P(\nu |\mathcal{E}_A, \mathcal{E}_B) = 
K_{n,n/2} \, g(\nu,\mathcal{E}_A) \, g(\nu,\mathcal{E}_B) \prod_{j < k=1}^{n/2} ( \nu_j^2 - \nu_k^2 )^2 \prod_{l=1}^{n/2} \nu_l^2 \, ,
\end{equation}
where
\begin{equation}\label{eq:gdef}
g(\nu,\mathcal{E}_A) := K_{n/2} \hspace{-0.1cm} \int \delta\bigg( \mathcal{E}_A  -\frac{1}{2} \sum_{h,k=1}^{n/2} |{U_A}_{hk}|^2 \, {\lambda_A}_h \, \nu_k \bigg) \hspace{-0.1cm} \prod_{h < k =1}^{n/2} \hspace{-0.1cm} |{\lambda_A}_h-{\lambda_A}_k| \d\lambda_A \d\mu(U_A) \, ,
\end{equation}
and $g(\nu,\mathcal{E}_B)$ is given by the analogous expression.

Let us apply Eq.~\eqref{EConstraint} to some simple but enlightening cases.

\subsection{The case of $1+1$ mode}

For $n = 2$ and $n_A = n_B = 1$, the mean energies~\eqref{conA}-\eqref{conB} read
\begin{equation}
\mathcal{E}_A  = \frac{1}{2} \lambda_A \nu \, , \quad\mathcal{E}_B  =  \frac{1}{2} \lambda_B \nu \, . 
\end{equation}
Thus, the integral in~\eqref{eq:gdef} becomes
\begin{equation}
g(\nu,\mathcal{E}_A) = K_1 \int \delta\left(\mathcal{E}_A-\frac{1}{2} {\lambda_A}_1 \nu\right) \, \d{\lambda_A}_1 = \frac{2 K_1}{\nu} \, .
\end{equation}
Therefore, after fixing the values of the normalization constants $K_1$, $K_{2,1}$ the probability 
measure~\eqref{EConstraint} of the symplectic eigenvalues reads
\begin{equation}
P(\nu |{\mathcal{E}_A,\mathcal{E}_B})= \frac{1}{\min{\{ \mathcal{E}_A, \mathcal{E}_B \}}-1} \, , \qquad 
\nu \in [1,\min{\{ \mathcal{E}_A, \mathcal{E}_B \}}] \, .
\end{equation}

\subsection{The case of $2+2$ modes}

For $n = 4$ and $n_A = n_B = 2$, the probability measure of the symplectic eigenvalues reads
\begin{equation}
P(\nu_1,\nu_2  |{\mathcal{E}_A,\mathcal{E}_B}) 
= K_{4,2} \, g(\nu_1,\nu_2,\mathcal{E}_A) \, g(\nu_1,\nu_2,\mathcal{E}_B) \left(\nu_1^2-\nu_2^2\right)^2 \nu_1^2 \nu_2^2 \, ,
\end{equation}
with $g$ given by~(\ref{eq:gdef}).
The $2 \times 2$ unitary matrix acting on the subsystem $A$ can be parameterized as
\begin{equation}
U_A = \left( \begin{array}{cc}
\e^{\ii\varphi}\cos{(\vartheta/2)}  & \e^{\ii\chi}\sin{(\vartheta/2)} \\
-\e^{-\ii\chi}\sin{(\vartheta/2)} & \e^{-\ii\varphi}\cos{(\vartheta/2)}
\end{array} \right) \, ,
\end{equation}
with $\vartheta\in[0,\pi]$, $\varphi, \chi \in [0,2\pi]$, and the 
normalized Haar measure on $\mathrm{U}(2)$ reads 
\begin{equation}
\d\mu(U_A)=2^{-1}(2\pi)^{-2} \sin{\vartheta} \d\vartheta \d\varphi \d\chi \, .
\end{equation}
Then, from \eqref{conA}, \eqref{conB} it follows
\begin{equation}
\mathcal{E}_A = \frac{1}{4} \left( {\lambda_A}_1+{\lambda_A}_2 \right) \left( \nu_1+\nu_2 \right) + \frac{1}{4} \cos{\vartheta} \left( {\lambda_A}_1-{\lambda_A}_2 \right) \left( \nu_1-\nu_2 \right) \, .
\end{equation}
After integration over $\d\varphi \d\chi$ we get
\begin{eqnarray}
g(\nu_1,\nu_2, \mathcal{E}_A) 
& = & \frac{K_2}{2} \int \d{\lambda_A}_1 \, \d{\lambda_A}_2 \, \d\cos{\vartheta} \, |{\lambda_A}_1-{\lambda_A}_2| \nonumber \\
&   & \qquad \times \delta\left(\mathcal{E}_A-\frac{1}{4}\left({\lambda_A}_1+{\lambda_A}_2\right) \left(\nu_1+\nu_2\right) -\frac{1}{4} \cos{\vartheta} \left( {\lambda_A}_1-{\lambda_A}_2\right)\left(\nu_1-\nu_2\right) \right) \nonumber \\
& = & \frac{2 K_2}{|\nu_1-\nu_2|} \int_D \d{\lambda_A}_1 \, \d{\lambda_A}_2 \, . 
\end{eqnarray}
The domain $D$ for the $\lambda_A$ variables is determined by the inequality
\begin{equation}
\left| \frac{4\mathcal{E}_A - ({\lambda_A}_1+{\lambda_A}_2)(\nu_1+\nu_2)}{({\lambda_A}_1-{\lambda_A}_2)(\nu_1-\nu_2)} \right| \leq 1 \, ,
\end{equation}
yielding
\begin{equation}
\int_D \d{\lambda_A}_1 \, \d{\lambda_A}_2 = \frac{|\nu_1-\nu_2|\left[ 2\mathcal{E}_A - (\nu_1+\nu_2) \right]^2}{\nu_1 \nu_2 (\nu_1+\nu_2)} \, .
\end{equation}
Finally, we obtain
\begin{equation}
P(\nu_1,\nu_2  |{\mathcal{E}_A,\mathcal{E}_B})
=  4 \, K_{4,2} \, K_2^2 \,(\nu_1-\nu_2)^2 \left[ 2\mathcal{E}_A - (\nu_1+\nu_2) \right]^2 \left[ 2\mathcal{E}_B - (\nu_1+\nu_2) \right]^2 \, , 
\end{equation}
where the range of the symplectic eigenvalues is determined by the inequalities
\begin{equation}
\left\{\begin{array}{l}
\nu_1 \geq 1 \, ,\\
\nu_2 \geq 1 \, ,\\
\nu_1 + \nu_2 \leq 2 \min{\{ \mathcal{E}_A , \mathcal{E}_B \}} \, .
\end{array}\right.
\end{equation}
The probability measure is zero for $\nu_1=\nu_2$, and attains its maximum
when one of the two symplectic eigenvalues is equal to one.

\subsection{Nonlinear optical parametric processes}\label{Sec:subM}

The measure (\ref{invm}) has been obtained by requiring the invariance over the whole
manifold of Gaussian states $\mathcal{G}$. However, in physical applications, one is mostly  
interested in submanifolds of states arising from specific physical processes.
The most relevant ones in which multimode Gaussian states are produced
are nonlinear optical parametric processes, see e.g.\ Ref.~\onlinecite{Ch}.
The Hamiltonian governing these kind of physical processes induces a natural bipartition of the bosonic system into two 
disjoint subsets of canonical modes, commonly referred to as {\it signal} and {\it idler} modes, which breaks
the symmetry under permutation of the modes.

Let $\bar{\mathcal G}$ (with $\bar{\mathcal G}\subset\mathcal{G}$) be the submanifold of such $n$-mode Gaussian states
with a given bipartition in $n_A$ signal modes and $n_B = n - n_A (\geq n_A)$ idler modes.
The quadratic Hamiltonians generating them are of the form
\begin{equation}\label{H1}
H_{\bar{G}} = \theta + \sum_{h,k=1}^{n_A} {N_A}_{hk} {a}_h^\dag {a}_k + \sum_{h,k=1}^{n_B} {N_B}_{hk} {b}_h^\dag {b}_k 
+ \sum_{h=1}^{n_A} \sum_{k=1}^{n_B} \left( M_{hk} {a}_h {b}_k + M_{hk}^* {a}^\dag_h {b}^\dag_k \right) \, ,
\end{equation}
where $N_A$ and $N_B$ are Hermitian. 
Comparing with Eq.~(\ref{GUnitary}), the linear terms and some of the quadratic terms are dropped.
The subgroup generated by these Hamiltonians is a projective representation of $\mathrm{SU}(n_A,n_B)$~\cite{Puri}.
According to Ref.~\onlinecite{Perelomov}, the Gaussian states of the form $\psi_{\bar{G}} = \exp{(-\ii H_{\bar{G}})} |0\rangle$
are hence coherent states for $\mathrm{SU}(n_A,n_B)$.
Following Ref.~\onlinecite{Puri}, these states can be written as
\begin{equation}
\psi_{\bar{G}} = \e^{-\ii\theta} \exp{\bigg( \sum_{h=1}^{n_A} \sum_{k=1}^{n_B} F_{hk} \, {a}_h {b}_k - \sum_{h=1}^{n_A} \sum_{k=1}^{n_B} F_{hk}^* \, {a}_h^\dag {b}_k^\dag \bigg)} |0\rangle \, .
\end{equation}
The canonical form for this class of states can be obtained from the singular value
decomposition of the matrix $F$: 
\begin{equation}
F_{hk} = \sum_{j \leq n_A} U_{hj} r_{j} V_{kj} \, ,
\end{equation}
with
$U \in \mathrm{U}(n_A)$, $V \in \mathrm{U}(n_B)$, and $r_j \geq 0$ are its singular values. 
By virtue of Eq.~(\ref{JS-id}), we can find a pair of local Gaussian unitaries 
$\mathcal{U}^A_{\bar{G}}$, $\mathcal{U}^B_{\bar{G}}$ such that 
\begin{eqnarray} 
{\mathcal{U}^A_{\bar{G}}}^\dagger  \, {a}_h \, \mathcal{U}^A_{\bar{G}} = \sum_{j=1}^{n_A} U^*_{hj} \, {a}_j \, ,\quad
{\mathcal{U}^B_{\bar{G}}}^\dagger \, {b}_k \, \mathcal{U}^B_{\bar{G}} = \sum_{i=1}^{n_B} V^*_{ki} \, {b}_i \, .
\end{eqnarray}
Thus the canonical forms of these states read
\begin{equation}
\psi^c_{\bar{G}} = \mathcal{U}^A_{\bar{G}} \otimes \mathcal{U}^B_{\bar{G}} \, \psi_{\bar{G}} = \prod_{k=1}^{n_A} \exp{\left( r_k {a}_k {b}_k - r_k {a}_k^\dag {b}_k^\dag \right)} |0\rangle \, ,
\end{equation}
where the local Gaussian unitaries have the form
\begin{eqnarray}
\mathcal{U}^A_{\bar{G}} & = & \exp{\bigg( -\ii \sum_{h,k=1}^{n_A} {T_A}_{hk} \, {a}^\dag_h {a}_k \bigg)} \, , \label{GUA} \\
\mathcal{U}^B_{\bar{G}} & = & \exp{\bigg( -\ii \sum_{h,k=1}^{n_B} {T_B}_{hk} \, {b}^\dag_h {b}_k \bigg)} \, , \label{GUB}
\end{eqnarray}
with $T_A$, $T_B$ Hermitian matrices.

Finally, the following Theorem holds:
\begin{theorem}\label{theoremmeasure3}
The invariant measure on the submanifold $\bar{\mathcal G}$ has the following expression
\begin{equation}\label{PDCmeasure}
\d\mu_{\bar{G}} = \bar{K}_{n,n_A} \prod_{h<k=1}^{n_A} \left( \nu_h - \nu_k \right)^2 \prod_{j=1}^{n_A} \left( \nu_j - 1 \right)^{n_B-n_A} \d\nu \, \d\mu(\bar{\alpha}_A) \, \d\mu(\bar{\alpha}_B) \, \d\theta \, ,
\end{equation}
where $\bar{K}_{n,n_A}$ is a normalization constant and the local degrees of freedom 
$\bar{\alpha}_A$, $\bar{\alpha}_B$, are induced by the local unitary transformations 
$\mathcal{U}^A_{\bar{G}}$, $\mathcal{U}^B_{\bar{G}}$.
\end{theorem}
The proof is given in SubSection \ref{pT3}.

The expression for the mean energy of the subsystems takes a particular simple form on
this submanifold, see Section \ref{App-energy}. 
For instance, for a balanced bipartition, $n_A=n_B=n/2$ ($n$ even),
\begin{equation}
\mathcal{E}_A = \mathcal{E}_B = \frac{1}{2} \sum_{k=1}^{n/2} \nu_k \, .
\end{equation}
The corresponding probability measure of the symplectic eigenvalues at fixed mean energy is 
\begin{equation}
\d\mu(\nu | \mathcal{E}) = \bar{K}_{n,n/2} \, \delta\bigg( \mathcal{E} - \frac{1}{2} \sum_{j=1}^{n/2} \nu_j \bigg) \prod_{h<k=1}^{n/2} \left( \nu_h - \nu_k \right)^2  \, \d\nu \, ,
\end{equation}
with $\mathcal{E}=\mathcal{E}_A=\mathcal{E}_B$.

\section{Proofs of the Theorems}\label{sec:proofs}

\subsection{Proof of Theorem \ref{th1}}\label{proof}

In order to derive the invariant measure~\eqref{invm} we consider the manifold of the
Gaussian pure states.
Such a manifold is embedded in the $n$-mode Hilbert space, $L^2(\mathbb{R}^n)$, which in turn is 
considered as a real vector space endowed with the real scalar product induced
by the standard Hermitian scalar product.

As a preliminary remark, notice that 
there is a common dense set of analytic vectors, containing $n$-mode numbers states~(\ref{eq:Hermite}),  
for the quadratic Hamiltonians $H_G$. That is,  any vector $\psi$ in this set satisfies:  $\psi\in \bigcap_n D(H_G^n)$, 
with $D(H_G^n)$  the domain of  $H_G^n$,
and $\sum_n \|H_G^n \psi \| t^n/n! < \infty$ for some $t>0$ \cite{ReedSimon}. 
This subspace is left invariant by the action of the Gaussian unitaries, generated by the quadratic Hamiltonians. 
Thus, on the manifold $\mathcal{G}$ of Gaussian pure states, which is the orbit of the vacuum state, the action of 
the Gaussian unitary group is smooth and can be derived at will.

Therefore, let us consider a system $\{ H_\alpha \}_\alpha$ of generators of the algebra 
of the group of Gaussian unitary transformations, and evaluate the corresponding
infinitesimal transformations on the canonical state: 
\begin{eqnarray}
\e^{-\ii \, \d\alpha H_\alpha} \psi_G^c -  \psi_G^c & = & -\ii \, \d\alpha\, H_\alpha \psi_G^c =  -\ii\, \d\alpha\,   U_{\bm{r}} \left( U_{\bm{r}}^\dag H_\alpha U_{\bm{r}} \right) |0\rangle \nonumber\\
& =: &  -\ii \, \d\alpha \, U_{\bm{r}} {\psi_G^c}_\alpha \, ,
\end{eqnarray}
where 
\begin{eqnarray}
&&U_{\bm{r}} = \prod_{k=1}^{n_A} \exp{\left[r_k ({a}_k {b}_k -   {a}_k^\dag {b}_k^\dag) \right]} \, , \\
&&{\psi_G^c}_\alpha= U_{\bm{r}}^\dag H_\alpha U_{\bm{r}}  |0\rangle \, ,
\end{eqnarray}
and $\bm{r}=(r_1, r_2, \dots, r_{n_A})\in \mathbb{R}^{n_A}$. 
Therefore, by neglecting the unitary factor $-\ii U_{\bm{r}}$,  we can define a set of vector-valued one-forms $\{ \d\alpha \,{\psi_G^c}_\alpha \}_\alpha$.
By construction, the volume form generated by a maximal subset of linearly 
independent one-forms is the desired invariant measure on the manifold of
Gaussian states.  We then proceed as follows.
By looking at the Hilbert space as a real vector space, we fix a 
suitable system $\{ \zeta_\beta \}_\beta$, orthonormal with respect to the 
real scalar product.
We hence expand the one-forms in terms of this set, i.e.,
\begin{equation}
{\psi^c_G}_\alpha = \sum_{\beta} \, J_{\alpha\beta} \, \zeta_\beta \, .
\end{equation}
Finally, the invariant measure is given in terms of the determinant
of the matrix of coefficients:
\begin{equation}
\d\mu_G = |\det{J}| \wedge_\alpha \d\alpha \, .
\end{equation}
 In order to write $\d \mu_G$ as a product of a measure
on the symplectic eigenvalues and a measure on the local degrees of freedom, 
we consider a basis for the algebra of the group of Gaussian unitary transformations composed by: 
\begin{itemize}
\item the generator of the scalar-phase shift;
\item a basis for the generators which are linear in the canonical operators;
\item a basis for the quadratic generators acting on subsystem $A$;
\item a basis for the quadratic generators acting on subsystem $B$; 
\item a suitable set of linearly independent nonlocal generators responsible for the variations of the symplectic eigenvalues.
\end{itemize}
Before going into the details of the calculations, let us have a closer 
look at the action of the unitary transformation $U_{\bm{r}}$ on the 
canonical operators. The following identities hold:
\begin{itemize}
\item for $k \leq n_A$
\begin{eqnarray}
U_{\bm{r}}^\dag {a}_k U_{\bm{r}} & = & \cosh{r_k} \, {a}_k + \sinh{r_k} \, {b}_k^\dag \, , \label{A} \\ 
U_{\bm{r}}^\dag {b}_k U_{\bm{r}} & = & \cosh{r_k} \, {b}_k + \sinh{r_k} \, {a}_k^\dag \, , \label{dB1}
\end{eqnarray}
\item for $k >n_A$
\begin{equation}\label{B2}
U_{\bm{r}}^\dag {b}_k U_{\bm{r}} = {b}_k \, . 
\end{equation}
\end{itemize}

The Lie algebra of the group of Gaussian unitaries is composed of Hamiltonians which
are (at most quadratic) functions of the canonical operators, 
\begin{equation}
H_\alpha = H_\alpha(a,a^\dag) \, ,
\end{equation}
where we have introduced the short-hand notation
\begin{eqnarray}
a      & = & ({a}_1 , \dots , {a}_{n_A},{b}_1 , \dots , {b}_{n_B}) \, , \\
a^\dag & = & ({a}^\dag_1 , \dots , {a}^\dag_{n_A},{b}^\dag_1 , \dots , {b}^\dag_{n_B}) \, . 
\end{eqnarray}
Below we compute the corresponding vector-valued one-forms using the following relation
\begin{equation}
\d\alpha\, {\psi^c_G}_\alpha  =  \d\alpha \left( U^\dag_{\bm{r}} H_\alpha(a,a^\dag) U_{\bm{r}} \right) |0\rangle =  \d\alpha\, H_\alpha( a' , {a'}^\dag ) |0\rangle \, ,
\end{equation}
where $a'= U^\dag_{\bm{r}} a U_{\bm{r}}$, ${a'}^\dag = U^\dag_{\bm{r}} a^\dag U_{\bm{r}}$
are explicitly given by Eqs.~(\ref{A})-(\ref{B2}).
It follows that the vector-valued one-forms belong to the linear span of the following
vectors:
\begin{itemize}
\item for $k \leq n_A$
\begin{eqnarray}
|k,0,0\rangle &:=& {a}_k^\dag |0\rangle \, , \\
|0,k,0\rangle &:=& {b}_k^\dag |0\rangle \, , \\
|k_2,0,0\rangle &:=& 2^{-1/2}({a}_k^\dag)^2 |0\rangle \, ,\\
|0,k_2,0\rangle &:=& 2^{-1/2}({b}_k^\dag)^2 |0\rangle \, , 
\end{eqnarray}
\item for $k>n_A$
\begin{equation}
|0,0,k\rangle := {b}_k^\dag |0\rangle \, ,
\end{equation}
\end{itemize}
\begin{itemize}
\item for $h \leq n_A$ , $k \leq n_A$
\begin{eqnarray}
|hk,0,0\rangle &:=& {a}_h^\dag {a}_k^\dag |0\rangle \, , \\
|0,hk,0\rangle &:=& {b}_h^\dag {b}_k^\dag |0\rangle \, , \\
|h,k,0\rangle &:=& {a}_h^\dag {b}_k^\dag |0\rangle \, , 
\end{eqnarray}
\item $h \leq n_A$, $k>n_A$
\begin{eqnarray}
|h,0,k\rangle &:=& {a}_h^\dag {b}_{k}^\dag |0\rangle \, , \\
|0,h,k\rangle &:=& {b}_h^\dag {b}_{k}^\dag |0\rangle \, . 
\end{eqnarray}
\end{itemize}

In the following we compute the explicit expressions of the one-forms.

\subsubsection{The scalar term}

By variation of the parameter $\theta$, corresponding to a scalar phase-factor,
we obtain the one-form
\begin{equation}\label{gps}
\d\theta |0\rangle \, .
\end{equation}

\subsubsection{The linear terms}
Let us now consider the linear generators, proportional to the complex vector $\xi$.
We consider variations of the real parameters
\begin{eqnarray}
&& \{ \mathrm{Re}({\xi_A}_k) , \mathrm{Im}({\xi_A}_k) \}_{k \leq n_A} \, , \\
&& \{ \mathrm{Re}({\xi_B}_k) , \mathrm{Im}({\xi_B}_k) \}_{k \leq  n_A} \, ,\\ 
&& \{ \mathrm{Re}({\xi_B}_k) , \mathrm{Im}({\xi_B}_k) \}_{n_A<k \leq n_B} \, , 
\end{eqnarray}
where $\mathrm{Re}$ and $\mathrm{Im}$ denote the real and the imaginary part.
Thus, using the identities~(\ref{A})-(\ref{B2}), we get the following expressions
for the corresponding one-forms:
\begin{itemize}
\item for $k \leq n_A$
\begin{eqnarray}
&& \d\mathrm{Re}({\xi_A}_k) \, \left( \cosh{r_k} |k,0,0\rangle + \sinh{r_k} |0,k,0\rangle \right) \, , \label{1p} \\
&& \ii \, \d\mathrm{Im}({\xi_A}_k) \, \left( \cosh{r_k} |k,0,0\rangle + \sinh{r_k} |0,k,0\rangle \right) \, , \\
&& \d\mathrm{Re}({\xi_B}_k) \, \left( \sinh{r_k} |k,0,0\rangle + \cosh{r_k} |0,k,0\rangle \right) \, , \\
&& \ii \, \d\mathrm{Im}({\xi_B}_k) \, \left( \sinh{r_k} |k,0,0\rangle + \cosh{r_k} |0,k,0\rangle \right) \, , 
\end{eqnarray}
\item for $k > n_A$
\begin{eqnarray}
&& \d\mathrm{Re}({\xi_B}_k) \, |0,0,k\rangle \, , \\
&& \ii \, \d\mathrm{Im}({\xi_B}_k) \, |0,0,k\rangle \, . \label{1u}
\end{eqnarray}
\end{itemize}

\subsubsection{The quadratic terms: subsystem $A$}

We now consider the generators which are quadratic in the canonical operators
of subsystem $A$:
\begin{equation}
H_A = \sum_{h,k=1}^{n_A} {M_A}_{hk} \, {a}_h {a}_k + {M_A}^*_{hk} \, {a}_h^\dag {a}_k^\dag + {N_A}_{hk} \, {a}_h^\dag {a}_k \, ,
\end{equation}
where $M_A$ is a complex-valued symmetric matrix, and $N_A$ is a Hermitian matrix. 

Considering the variations of the parameters $\mathrm{Re}({M_A}_{hk})$, $\mathrm{Im}({M_A}_{hk})$,
for  $h \leq k \leq n_A$, we have the one-forms
\begin{eqnarray}
&& \d\mathrm{Re}({M_A}_{hk}) \, \left( \cosh{r_h} \cosh{r_k} |hk,0,0\rangle + \sinh{r_h} \sinh{r_k} |0,hk,0\rangle \right) \, , \label{MAp} \\
&& \d\mathrm{Re}({M_A}_{kk}) \, \sqrt{2} \left( \cosh^2{r_k} |k_2,0,0\rangle + \sinh^2{r_k} |0,k_2,0\rangle \right) \, , \\   
&& -\ii\, \d\mathrm{Im}({M_A}_{hk}) \, ( \cosh{r_h} \cosh{r_k} |hk,0,0\rangle - \sinh{r_h} \sinh{r_k} |0,hk,0\rangle ) \, , \\
&& -\ii\, \d\mathrm{Im}({M_A}_{kk}) \, \sqrt{2} \left( \cosh^2{r_k} |k_2,0,0\rangle - \sinh^2{r_k} |0,k_2,0\rangle \right) \, . \label{MAu}
\end{eqnarray}

The variations of the parameters $\mathrm{Re}({N_A}_{hk})$, $\mathrm{Im}({N_A}_{hk})$ for $h < k \leq n_A$, 
and ${N_A}_{kk}$ for $k \leq n_A$, yield
\begin{eqnarray}
&& \d\mathrm{Re}({N_A}_{hk}) \, \left( \cosh{r_h} \sinh{r_k} |h,k,0\rangle + \sinh{r_h} \cosh{r_k} |k,h,0\rangle \right) \, , \label{NAp} \\
&& \d{N_A}_{kk} \, \left( \cosh{r_k} \sinh{r_k} |k,k,0\rangle + \sinh^2{r_k} |0\rangle \right) \, , \label{dip1} \\
&& \ii \, \d\mathrm{Im}({N_A}_{hk}) \,\left( \cosh{r_h} \sinh{r_k} |h,k,0\rangle - \sinh{r_h} \cosh{r_k} |k,h,0\rangle \right) \, . \label{NAu}
\end{eqnarray}

\subsubsection{The quadratic terms: subsystem $B$}

Moving to the local transformations on the subsystem $B$, we first consider
the quadratic Hamiltonians involving the operators $\{ {b}_k , {b}_k^\dag \}_{k \leq n_A}$,
which are of the form
\begin{equation}
H_B = \sum_{h,k=1}^{n_A} {M_B}_{hk} \, {b}_h {b}_k + {M_B}^*_{hk} \, {b}_h^\dag {b}_k^\dag + {N_B}_{hk} \, {b}_h^\dag {b}_k \, .
\end{equation}
Proceeding as in the case of subsystem $A$, we obtain
the following one-forms from the variations of the matrix $M_B$:
\begin{eqnarray}
&& \d\mathrm{Re}({M_B}_{hk}) \, \left( \sinh{r_h} \sinh{r_k} |hk,0,0\rangle + \cosh{r_h} \cosh{r_k} |0,hk,0\rangle \right) \, , \label{MBp} \\
&& \d\mathrm{Re}({M_B}_{kk}) \, \sqrt{2} \left( \sinh^2{r_k} |k_2,0,0\rangle + \cosh^2{r_k} |0,k_2,0\rangle \right) \, , \\
&& \ii \, \d\mathrm{Im}({M_B}_{hk}) \,\left( \sinh{r_h} \sinh{r_k} |hk,0,0\rangle - \cosh{r_h} \cosh{r_k} |0,hk,0\rangle \right) \, , \\
&& \ii \, \d\mathrm{Im}({M_B}_{kk}) \, \sqrt{2} \left( \sinh^2{r_k} |k_2,0,0\rangle - \cosh^2{r_k} |0,k_2,0\rangle \right) \, . \label{MBu}
\end{eqnarray}
From the variations of the matrix $N_B$ we obtain the one-forms
\begin{eqnarray}
&& \d\mathrm{Re}({N_B}_{hk}) \, \left( \sinh{r_h} \cosh{r_k} |h,k,0\rangle + \cosh{r_h} \sinh{r_k} |k,h,0\rangle \right) \, , \label{NBp} \\
&& \d{N_B}_{kk} \, \left( \sinh{r_k} \cosh{r_k} |k,k,0\rangle + \sinh^2{r_k} |0\rangle \right) \, , \label{dip2} \\
&& -\ii \, \d\mathrm{Im}({N_B}_{hk}) \, \left( \sinh{r_h} \cosh{r_k} |h,k,0\rangle - \cosh{r_h} \sinh{r_k} |k,h,0\rangle \right) \, . \label{NBu}
\end{eqnarray}

Then we consider the Hamiltonians coupling the operators 
$\{ {b}_k , {b}_k^\dag \}_{k \leq n_A}$ with $\{ {b}_k , {b}_k^\dag \}_{k > n_A}$, that is,
\begin{equation}
H_{B}  =  \sum_{h=1}^{n_A} \sum_{k=n_A+1}^{n_B} \left( {P_{B'}}_{hk} \, {b}_h {b}_k + {P_{B'}^*}_{hk} \, {b}_h^\dag {b}_k^\dag + {Q_{B'}}_{hk} \, {b}_h {b}_k^\dag + {Q_{B'}^*}_{hk} \, {b}_h^\dag {b}_k \right) \, ,
\end{equation}
where $P_{B'}$ and $Q_{B'}$ are complex-valued matrices.
For $h \leq n_A$ and $n_A < k \leq n_B$ we get
\begin{eqnarray}
&& \d\mathrm{Re}({P_{B'}}_{hk}) \, \cosh{r_h} |0,h,k\rangle \, , \label{MB1p} \\
&& \ii \, \d\mathrm{Im}({P_{B'}}_{hk}) \, \cosh{r_h} |0,h,k\rangle \, , \label{MB1u} \\
&& \d\mathrm{Re}({Q_{B'}}_{hk}) \, \sinh{r_h} |h,0,k\rangle \, , \label{NB1p} \\
&& \ii \,\d\mathrm{Im}({Q_{B'}}_{hk}) \, \sinh{r_h} |h,0,k\rangle \, . \label{NB1u}
\end{eqnarray}

Finally, let us consider the quadratic Hamiltonians containing only the operators 
$\{ {b}_k , {b}_k^\dag \}_{k>n_A}$:
\begin{equation}
H_{B'} = \sum_{h,k=n_A+1}^{n_B} {M_{B'}}_{hk} \, {b}_h {b}_k + {M_{B'}}^*_{hk} \, {b}_h^\dag {b}_k^\dag + {N_{B'}}_{hk} \, {b}_h^\dag {b}_k \, ,
\end{equation}
where the matrix $M_{B'}$ is complex-valued and symmetric and the matrix $N_{B'}$ is Hermitian.
We obtain that the variations of the elements of the matrix $N_{B'}$
yield vanishing one-forms. 
The non-zero one-forms are generated by the variations of $M_{B'}$,
\begin{eqnarray}
\d\mathrm{Re}({M_{B'}}_{hk}) \, 2 |0,0,hk\rangle \, ,&& \label{3p} \\
\d\mathrm{Re}({M_{B'}}_{kk}) \, \sqrt{2} |0,0,k_2\rangle \, ,&&\\
-\ii \, \d\mathrm{Im}({M_{B'}}_{hk}) \, 2 |0,0,hk\rangle \, ,&&\\
-\ii \, \d\mathrm{Im}({M_{B'}}_{kk}) \, \sqrt{2} |0,0,k_2\rangle \label{3u}\, .&&
\end{eqnarray}

\subsubsection{The nonlocal generators}

It remains to consider the nonlocal transformations whose action changes the
symplectic eigenvalues. These are generated by Hamiltonians
of the form 
\begin{equation}
(H_{NL})_k = \ii \left( {a}_k {b}_k - {a}_k^\dag {b}_k^\dag \right) \, ,  \qquad k \leq n_A \, ,  
\end{equation}
The corresponding one-forms are
\begin{equation}\label{nonloc}
- \ii \, \d r_k |k,k,0\rangle =  \frac{-\ii \, \d\nu_k}{2\sinh{2r_k}} |k,k,0\rangle \, , \quad k \leq n_A \, .
\end{equation}

\subsubsection{The invariant measure}

We can now compute the invariant measure on the manifold of Gaussian states.

First, we consider the one-forms corresponding to linear Hamiltonians.
The matrix of coefficients is readily obtained from Eqs.~(\ref{1p})-(\ref{1u}),
from which we obtain the following factor in the invariant measure
\begin{equation}
\prod_{h=1}^{n_A} \d\mathrm{Re}({\xi_A}_h)\d\mathrm{Re}({\xi_A}_h) \prod_{k=1}^{n_B} \d\mathrm{Re}({\xi_B}_k)\d\mathrm{Re}({\xi_B}_k) = \d\xi_A \d\xi_B \, ,
\end{equation}

Second, we consider the one-forms corresponding to the scalar phase-shift and the quadratic Hamiltonians.
The matrix of coefficients can be straightforward obtained from
Eqs.~(\ref{gps}), (\ref{MAp})-(\ref{NAu}), (\ref{MBp})-(\ref{NBu}), (\ref{MB1p})-(\ref{NB1u}),
(\ref{3p})-(\ref{3u}), and (\ref{nonloc}). 
A maximal subset of linearly independent one-forms can be obtained by eliminating 
the one-form in (\ref{dip1}), which is proportional to the one in 
(\ref{dip2}) due to the symmetry of the canonical form (\ref{CVS}).

From the matrix of coefficient one gets
\begin{equation}
|\det{J}| = \mathcal{C} \, \prod_{h<k=1}^{n_A} \left( \nu_h^2 - \nu_k^2 \right)^2 \prod_{j=1}^{n_A} \nu_j^2 \left( \nu_j^2 - 1 \right)^{(n_B-n_A)} \, ,
\end{equation}
where $\mathcal{C}$ is a constant factor.
Finally, the invariant measure in Eq.~(\ref{invm}) is obtained by inserting the
differentials $\d\theta$, $\d\nu_1, \dots , \d\nu_{n_A}$, and identifying the
factors depending on the local degrees of freedom
\begin{eqnarray}
\d\mu(\alpha_A) & = & \d\xi_A \prod_{h<k=1}^{n_A} \d^2{M_A}_{hk} \d^2{N_A}_{hk} \prod_{i=1}^{n_A} \d^2{M_A}_{ii} \, ,  \\
\d\mu(\alpha_B) & = & \d\xi_B \prod_{h<k=1}^{n_B} \d^2{M_B}_{hk} \d^2{N_B}_{hk} \prod_{i=1}^{n_A} \d^2{M_B}_{ii} d{N_B}_{ii} \nonumber  \\
& & \times \prod_{j=1}^{n_A} \prod_{l=n_A+1}^{n_B} \d^2{P_{B'}}_{jl} \d^2{P_{B'}}_{jl} \d^2{Q_{B'}}_{jl} \d^2{Q_{B'}}_{jl} \prod_{p \leq q = n_A+1}^{n_B} \d^2{M_{B'}}_{pq} \, .
\end{eqnarray}

\subsection{Proof of Theorem \ref{theoremunitmeasure}} \label{measure2}

In order to derive the explicit expression of the Haar measure on the group of $n$-mode 
homogeneous Gaussian unitaries, we apply the unitary transformations on the vacuum state.
Using the Euler decomposition in Eq.~(\ref{gaussunitary2}) we obtain the $n$-mode 
homogeneous Gaussian state
\begin{equation}
\psi_G = \mathcal{U}_G |0\rangle = \e^{-\ii\theta}\exp{\bigg(-\ii \sum_{i,j=1}^n T_{ij} a_i^\dag a_j \bigg)} \exp{\bigg( \sum_{k=1}^n s_k a_k^2 - s_k (a_k^\dag)^2 \bigg)} |0\rangle \, ,
\end{equation}
where we have used $\exp{\left(-\ii \sum_{i,j=1}^n T'_{ij} a_i^\dag a_j\right)} |0\rangle = |0\rangle$.
Then, by variation of the parameter $\theta$, of the elements of $T$ and of the parameters $s=(s_1, s_2, \dots s_n)$, 
we obtain a set of vector valued one-forms.
By proceeding as in Section \ref{proof}, these one-forms can be used to derive an explicit expression
for the invariant measure on the manifold of homogeneous Gaussian states.
Then, the Haar measure on the group of homogeneous Gaussian unitaries can be readily obtained from the latter. 

By variation of the parameter $\theta$, we obtain the vector-valued one-form
\begin{equation}\label{form-theta}
\d\theta \, |0\rangle \, .
\end{equation}
The variations of the parameters $s$ yield
\begin{equation}\label{form-s}
\ii \, \d s_k \, \sqrt{2} |k_2\rangle \, , 
\end{equation}
where $|k_2\rangle = 2^{-1/2} ( a_k^\dag )^2 |0\rangle$. 
The variations of the parameters $T_{kk}$ yield the one-forms
\begin{equation}\label{form-Nkk}
\d T_{kk} \, \left[ 2^{-1/2} \sinh{2s_k} |k_2\rangle + (\sinh{s_k})^2 |0\rangle \right] \, . 
\end{equation}
Similarly, by variations of the parameters $\mathrm{Re}(T_{hk})$, we obtain
\begin{equation}\label{form-ReN}
\d\mathrm{Re}(T_{hk}) \, \sinh{(s_h+s_k)} |hk\rangle \, , \qquad  h < k \leq  n \, , 
\end{equation}
and the variations of the parameters $\mathrm{Im}(T_{hk})$ yield
\begin{equation}\label{form-ImN}
\ii \, \d\mathrm{Im}(T_{hk}) \, \sinh{(s_k-s_h)} |hk\rangle \, , \qquad h < k \leq n \, ,  
\end{equation}
where $|hk\rangle = a_h^\dag a_k^\dag |0\rangle$.

The volume form generated by these one-forms is by construction the invariant
measure on the considered submanifold of Gaussian states. 
First, we notice that the volume forms generated by the variations of the matrix 
elements of $T$ and $T'$ are
\begin{eqnarray}
\prod_{k=1}^n \d T_{kk} \prod_{i<j=1}^{n} \d^2T_{ij} & = & \d\mu(U) \, ,\\
\prod_{k=1}^n \d T'_{kk} \prod_{i<j=1}^{n} \d^2T'_{ij} & = & \d\mu(U') \, ,
\end{eqnarray}
where $\d\mu(U)$ and $\d\mu(U')$, with $U = \exp(-\ii T)$ and $U' = \exp(-\ii T')$,
denote the Haar measure on the unitary group $\mathrm{U}(n)$. 
Then, from the one-forms (\ref{form-theta})-(\ref{form-ImN}), we  derive the following expression for the Haar measure on the 
group of $n$-mode homogeneous Gaussian unitaries:
\begin{equation}
\d\mu(\mathcal{U}_G) = K_n \prod_{h<k=1}^{n} | \lambda_h - \lambda_k | \prod_{j=1}^n d\lambda_j \, \d\mu(U) \d\mu(U') \, ,
\end{equation}
with $\lambda_k=\cosh{2s_k}$, and $K_n$ a normalization factor.

\subsection{Proof of Lemma \ref{theoremmeanen}}\label{App-energy}

For a given $n$-mode homogeneous Gaussian state $\psi_G$, and a bipartition
of the system defined by two disjoint sets of canonical operators
\begin{equation}
\{ {a}_k , {a}_k^\dag \}_{k=1,\dots n_A}, \quad \{ {b}_k , {b}_k^\dag \}_{k=1,\dots n_B} \, ,
\end{equation}
$n_A + n_B = n$, we consider the mean value of the energy of one of the two subsystems.
The non-homogeneous case can be analyzed in a similar way, and gives rise to an additional 
term in the mean energy.
To fix the ideas we consider the mean energy of subsystem $A$,
\begin{equation}
\mathcal{E}_A = \frac{1}{2} \sum_{k=1}^{n_A} \langle \psi_G | ( {a}_k^\dag {a}_k + {a}_k {a}_k^\dag ) \psi_G\rangle \, ,
\end{equation}
and restrict to the case of balanced bipartition, $n_A = n_B = n/2$.

Using the canonical form (\ref{CVS}) we get
\begin{eqnarray}
\mathcal{E}_A & = & \frac{1}{2} \sum_{k=1}^{n/2} \langle \psi_G^c | {\mathcal{U}^A_G}^\dag ( {a}^\dag_k {a}_k + {a}_k {a}^\dag_k ) {\mathcal{U}^A_G} \psi_G^c \rangle \nonumber \\
& = & \frac{1}{2} \sum_{k=1}^{n/2} \langle \psi_G^c | ( a'^\dag_k {a'}_k + {a'}_k a'^\dag_k ) \psi_G^c \rangle \, ,
\end{eqnarray}
where ${a'}_k = {\mathcal{U}^A_G}^\dag {a}_k {\mathcal{U}^A_G}$ are $a'^\dag_k = {\mathcal{U}^A_G}^\dag {a}_k^\dag {\mathcal{U}^A_G}$
are linear combinations of the operators $\{ {a}_k , {a}_k^\dag \}_{k=1,\dots m}$.
The explicit form of $\mathcal{U}^A_G$ can be written starting from the Euler decomposition of the homogeneous 
Gaussian unitary, see Eq.~(\ref{gaussunitary2}), 
\begin{equation}
\mathcal{U}^A_G = \e^{-\ii \theta} \exp{\bigg( -\ii \sum_{i,j=1}^{n/2} {T_A}_{ij} {a}_i^\dag {a}_j \bigg)} \exp{\bigg( \sum_{k=1}^{n/2} {s_A}_k {a}_k^2 - {s_A}_k ({a}_k^\dag)^2 \bigg)}
 \exp{\bigg( -\ii \sum_{i,j=1}^{n/2} {T_A}'_{ij} {a}_i^\dag {a}_j \bigg)} \, . 
\end{equation}
Then, using Eqs.~(\ref{JS-id}), (\ref{localsq}) we get
\begin{equation}
\mathcal{E}_A = \frac{1}{2} \sum_{h,k=1}^{n/2} |{U_A}_{hk}|^2 \, {\lambda_A}_h \, \nu_k \, ,
\end{equation}
where $U_A = \exp{\left( -\ii T_A \right)}$, ${\lambda_A}_k = \cosh{2{s_A}_k}$, and $\nu_k$'s 
are the symplectic eigenvalues.
The analogous expression is obtained for the mean energy of subsystem $B$,
\begin{equation}
\mathcal{E}_B = \frac{1}{2} \sum_{h,k=1}^{n/2} |{U_B}_{hk}|^2 \, {\lambda_B}_h \, \nu_k \, .
\end{equation}

The local mean energies take a particular simple form for the submanifold of states
considered in Section~\ref{Sec:subM}.
In that case, from Eqs.~(\ref{GUA}), (\ref{GUB}) we get ${\lambda_A}_k = {\lambda_B}_k = 1$, 
which in turn implies
\begin{equation}
\mathcal{E}_A = \mathcal{E}_B = \frac{1}{2} \sum_{k=1}^{n/2} \nu_k \, .
\end{equation}

\subsection{Proof of Theorem \ref{theoremmeasure3}}\label{pT3}

To derive the invariant measure in Eq.~(\ref{PDCmeasure}), we proceed along the same steps of 
Section \ref{proof}, with the difference that only the terms which are compatible 
with the form of the Hamiltonian in Eq.~(\ref{H1}), and hence compatible with 
the local Gaussian unitaries in Eqs.~(\ref{GUA}), (\ref{GUB}), have to be retained.
Thus, the invariant measure is generated by the one-forms (\ref{gps}), (\ref{NAp})-(\ref{NAu}),
(\ref{NBp})-(\ref{NBu}), (\ref{NB1p})-(\ref{NB1u}), and (\ref{nonloc}).
This yields the expression in Eq.~(\ref{PDCmeasure}) where the factors depending on the local degrees
of freedom are explicitly given by
\begin{eqnarray}
\d\mu(\bar{\alpha}_A) & = & \prod_{h<k=1}^{n_A} \d^2{N_A}_{hk} \, , \\
\d\mu(\bar{\alpha}_B) & = & \prod_{h<k=1}^{n_B} \d^2{N_B}_{hk} \prod_{i=1}^{n_A} d{N_B}_{ii} \prod_{j=1}^{n_A} \prod_{l=n_A+1}^{n_B} \d^2{Q_{B'}}_{jl} \d^2{Q_{B'}}_{jl} \, . 
\end{eqnarray}

\begin{acknowledgments}
The work of CL and SM is supported by EU through the FET-Open Project HIP (FP7-ICT-221899). 
PF and GF acknowledge support by the University of Bari through the Project IDEA. 
GF acknowledges support by Istituto Nazionale di Alta Matematica and Gruppo Nazionale per la Fisica Matematica through the Project Giovani GNFM.
ADP was partially supported by the Italian Ministry of University and Research 
through FIRB-IDEAS Project no.\ RBID08B3FM.
\end{acknowledgments}

\end{document}